\documentclass[iop]{emulateapj}

\shorttitle{Classification of Stellar Orbits in Axisymmetric Galaxies}
\shortauthors{Li Holley-Bockelmann and Khan}

\begin{document}

\title{Classification of Stellar Orbits in Axisymmetric Galaxies}

\author{Baile Li\altaffilmark{1}, Kelly Holley-Bockelmann\altaffilmark{1,2}, and Fazeel Khan\altaffilmark{3,4,5} }
\affil{
\altaffilmark{1}{Department of Physics and Astronomy, Vanderbilt University,
    Nashville, TN 37235, USA; baile.li@vanderbilt.edu}\\     
\altaffilmark{2}{Department of Physics, Fisk University, Nashville, TN 37208, USA; k.holley@vanderbilt.edu}\\
\altaffilmark{3}{Department of Space Science, Institute of Space Technology, P.O. Box 2750 Islamabad, Pakistan; khan@ari.uni-heidelberg.de}\\
\altaffilmark{4}{Department of Physics, Government College University (GCU), 54000 Lahore, Pakistan}\\
\altaffilmark{5}{Astronomisches Rechen-Institut, Zentrum f\"ur Astronomie, University of Heidelberg, M\"onchhof-Strasse 12-14, D-69120 Heidelberg, Germany}
}

\begin{abstract}
It is known that two supermassive black holes (SMBHs) cannot merge in a spherical galaxy within a Hubble time; an emerging picture is that galaxy geometry, rotation, and large potential perturbations may usher the SMBH binary through the critical three-body scattering phase and ultimately drive the SMBH to coalesce. We explore the orbital content within an N-body model of a mildly-flattened, non-rotating, SMBH-embedded elliptical galaxy. When used as the foundation for a study on the SMBH binary coalescence, the black holes bypassed the binary stalling often seen within spherical galaxies and merged on Gyr timescales \citep{KHBJ13}.Using both frequency-mapping and angular momentum criteria, we identify a wealth of resonant orbits in the axisymmetric model, including saucers, that are absent from an otherwise identical spherical system and that can potentially interact with the binary. We quantified the set of orbits that could be scattered by the SMBH binary, and found that the axisymmetric model contained nearly seven times the number of these potential loss cone orbits compared to our equivalent spherical model. In this flattened model, the mass of these orbits is roughly 3 times of that of the SMBH, which is consistent with what the SMBH binary needs to scatter to transition into the gravitational wave regime.

\end{abstract}

\section{Introduction}

In Nature, a perfectly smooth and spherical galaxy is extremely rare -- and arguably may not exist at all. Nearly every galaxy has some degree of non-sphericity, be it axisymmetry, triaxiality, warps, or flares, and it is often the case that the galaxy shape varies with radius. The global shape of the galaxy potential, however, governs the motions of the stars and dark matter throughout. 

Galaxies can be grouped according to their shape as spherical, axisymmetric or triaxial. As the degree of symmetry in a galaxy decreases, there is more freedom in the orbit because there are fewer formal isolating integrals of motion. In a triaxial galaxy, for example, orbits do not have to conserve angular momentum, which admits a rich variety of regular resonant orbits \citep{NS83, GB85, MT99, MP04, MV11}, such as bananas, pretzels and boxes,  that can veer close to the supermassive black hole (SMBH) at the galactic center. The growth of SMBH can change the shape of a galaxy from triaxial to axisymmetric \citep{GB85,NMA85,MQ98,WF98,VM98, HB02}. SMBH can keep the shape of axisymmetric galaxies by inducing chaos and constraining the shape of regular orbits \citep{PM01}. 

The orbital content of a galaxy is important because it is the skeleton that defines its shape, structure, and evolution with time. In fact, it is believed that a solution to how black holes merge together and grow may lie with stellar orbits. Theory suggests that when galaxies merge, their two SMBHs sink to the center of the remnant and form a binary, whose orbit slowly shrinks by scattering stars away, but early simulations of the process show that the binary’s orbit stalls before the black holes plunge toward merger. This is ``the final parsec problem" \citep{MM03}, which has been solved recently by properly simulating SMBH binary evolution in galaxy mergers\citep{KJM11, PBBS11}, triaxial models \citep{BMSB06, HS06}, and most recently in an axisymmetric galaxy\citep{KHBJ13}. However, \citet{VAM14} argued that the rates of binary hardening within their own axisymmetric model highly depend on N, the number of particles in the simulations, in the range of $10^{5}\le N \le 10^{6}$. While \citet{KHBJ13} find binary hardening rates consistent with a full loss cone, \citet{VAM14} argue that their own models are far from the full loss cone regime and that the apparent binary evolution is dominated by collisional processes set by numerical relaxation. The apparent disagreement between these axisymmetric results may indicate that interpreting the coalescence time of SMBH binaries within N-body simulations of this sort must be done in conjunction with an analysis of the underlying orbit structure of the model. In this paper, we analyzed the orbit content of an N-body generated black hole embedded axisymmetric galaxy model \citep{KHBJ13} to understand which orbits could enable the binary black holes to pass through the final parsec to the gravitational radiation regime.

We focused on a census of the stars with pericenters well within the radius of influence of the SMBH, though we also take stock of the resonant orbits that populate the model in general as well. The paper is organized as follows. Section 2 describes our simulation method. Section 3 presents our results. We conclude with a discussion and conclusion in section 4. 

\section{Method}

We begin with a spherical galaxy model with a Hernquist density profile \citep{H90}, populated with $10^{6}$ equal-mass collisionless particles and a supermassive black hole of mass 0.005 fixed at the center. Then we “adiabatically squeeze” \citep{HB01} this spherical galaxy to generate an axisymmetric model with axis ratios $\frac{b}{a}=1$,$\frac{c}{a}=0.75$. This model was used in \citet{KHBJ13} as the background galaxy to study black hole binary coalescence. 

We construct our model with the galactic center in broadly in mind, so the mass of the SMBH in system units, 0.005, maps to $4\times 10^{6} M_{\odot}$. To pin down the length scale, we find the radius in the model where the enclosed stellar mass is roughly twice of that of the SMBH; in system units this radius is 0.05, while in the Milky Way, this radius is roughly 1 parsec \citep{GP00, SEA07, Ghez08, OKF09}. Given the mass and length scaling, the system unit velocities should be scaled by $\sim 450$ km/s and the system time can by scaled by  $\sim 4\times 10^{4}$ years. The highest velocity particle is only 1 $\%$ the speed of light, so we do not apply post-Newtonian corrections in our simulation. 

In general, the technique of orbital analysis involves following the particles within a fixed background galaxy potential. Ideally, the galaxy potential should be as smooth as possible to mitigate numerically-induced diffusion in the particle trajectories; this two-body relaxation will artificially induce chaotic orbit and can scatter particles out of resonant orbits \citep{HW92, K95, S03, HWK05, WK07a,WK07b}. To obtain a smoother potential we '8-fold' the model, reflecting each particle position about the principle axes \citep{HB01}. Further, we use a self consistent field (SCF) code\citep{HO92} to evolve the orbits in all six simulation series, which will be discussed in detail in the following. The SCF code is a particle-field code, where the particles do not interact with each other directly, but are accelerated by the global potential of the black hole-embedded galaxy. Here, the stellar potential and density are expressed as series expansion of ultraspherical harmonic functions. Here we employ nmax=10, lmax=6, though the results are largely unchanged for nmax=2-20, lmax=0-15.

We run each orbit for 100 dynamical times of an circular orbit with the same energy within the combined fixed potential from the supermassive black hole and the axisymmetric stellar model\citep{CA98}. We adjust the time step of each particle to ensure that fractional energy loss from integration errors is less than $10^{-7}$ for each orbit. Typical fractional energy loss is less than $10^{-11}$ over a time span much larger than Hubble time when the model is scaled to the Milky Way.

To analyze each orbit, we evenly sample the positions and Fourier transform the trajectory to obtain the principle frequencies that characterize the motion of the particle with respect to the x, y, and z axes. We can classify the orbit type according to the frequency ratio fx/fz and fy/fz \citep{L93,CA98}.  To distinguish a chaotic orbit from a stable one, we analyze the orbit in two time slices of 50 dynamical times; the frequency ratio of a chaotic orbit will vary between the two time slices.  To record the full information of the orbit, we also keep track of the pericenter distance, the minimum angular momentum and the minimum of the z component of the angular momentum for each particle. The resonances are marked by (u,v,w), which correspond to integers and are coefficients of the equation $u\cdot fx+v\cdot fy+w\cdot fz=0$.

To fully map the orbital structure of this potential, we conduct 3 experiments, and each experiment is comprised of the axisymmetric model and its spherical counterpart. The ``Galaxy" series simply analyzes the orbits of the particles directly within the original spherical N-body model and the final adiabatically-squeezed flat model. The advantage of the Galaxy set is that it directly probes the orbits that could eventually interact with the binary black hole in the \citet{KHBJ13} N-body simulation; since we use the axisymmetric model that results in a successful binary black hole coalescence, it is important to take stock of the orbits within. The disadvantage of this set is that it is merely one realization of the potential, and since a galaxy model is constructed from stars over a continuum of energies, it is difficult to compare our results to orbit analyses in the literature, where it is traditional to map out the orbital structure at a fixed energy. For this reason, we run ``3D" and ``2D" models that sample the phase space much more finely within 8 fixed energy slices. The ``2D'' series only maps orbits within the x-z plane, but this allows us a straightforward visualization of the resonant orbits, and allows us to construct a meaningful surface of section as well. See table 1 for more detail on each run.

For the 8 energy slices of the ``3D" and ``2D" models, the energy E=-2.5, -2, -1, -0.5, -0.4, -0.3, -0.2, -0.1, respectively. The stellar mass of particles with energy less than each E in the axisymmetric galaxy is respectively $1\times 10^{-4}$, $6\times 10^{-4}$, $10\%$, $40\%$, $45\%$, $55\%$, $70\%$, $80\%$ of the total stellar mass. The corresponding radius for a particle to run on a circular orbit with each E in the axisymmetric galaxy is respectively 0.0015, 0.0025, 0.025, 0.45, 0.7, 1, 2, 4. The units shown in all figures are model units unless otherwise indicated.

\begin{figure}[]
\centering
\includegraphics[width=75mm]{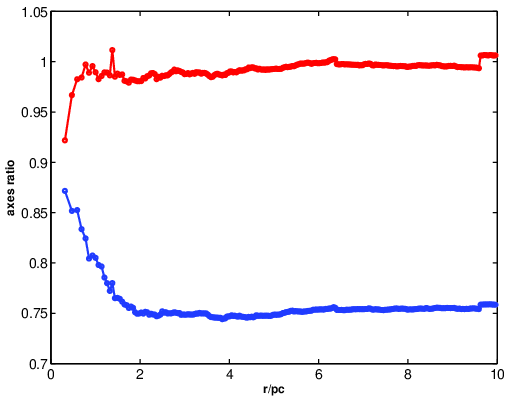}
\caption{Axes ratio b/a (red) and c/a (blue) in the inner 10 parsecs of the axisymmetric model. Though not plotted, the axis ratios are stable and the system is axisymmetric within 100 parsecs; at larger distances, the system becomes more spherical because the timescale for the applied velocity drag is non-adiabatic compared to typical orbital timescales there.  } 
\label{axes}
\end{figure}

Since this model was constructed non-analytically by dragging particles in velocity using an N-body simulation, it is not guaranteed to be a precisely homologous figure. We characterize the global shape by the axes ratios at the half mass radius, $b/a=1$ and $c/a=0.75$, and as can be seen in Figure \ref{axes}, the shape is fairly stable throughout the bulk of the model. The one big exception is at large radius, where the orbital time of the particles is so long that the squeezing technique is non-adiabatic and therefore the orbits of these outermost particles are largely unaffected by the applied velocity drag; this affects about 20$\%$ of the particles on the outer edge of the system. We should therefore expect that the orbital content of the outskirts of this galaxy model should mimic the spherical model and that we are missing the axisymmetric orbit families that would lie out there if the model were perfectly homologous. 

We note one other seemingly small detail: at the innermost part of the model, within the central 0.5 parsec, which is within the radius of influence of the SMBH, c/a is less flattened, around 0.85, and b/a trends below 1.0, around 0.96. Here, the model is actually triaxial with T=0.28. The mass fraction involved in the triaxial portion is small -- less than $ 0.25\%$ -- about half the SMBH mass. The finding of a technically triaxial shape inside the radius of influence of the SMBH may seem counter to previous work \citep{VM98, HB02}, which finds that the presence of a SMBH will act to sphericalize a triaxial shape. However, our finding is not inconsistent for several reasons. First, our model was embedded with a SMBH in place at its full mass before we morphed the galaxy shape, while most previous work focused on how galaxy models adjusts to a SMBH that starts at zero mass and grows. Second, earlier work may only have noted the trend toward a spherical figure (which we are in fact seeing -- note the axis ratios are both trending toward 0.9); they may have counted such minor triaxiality that we observe as essentially spherical. Finally, previous work followed the evolution of triaxial models over hundreds of dynamical times, while it is not clear how long the figure shape we observe will persist. 

\begin{table*}[]
\caption{model detail}
  \centering
  \begin{tabular}{c c c c c}
\hline
model name  & particles' initial condition & potential & model dimension & number of particles\\ \hline
Galaxy & axisymmetric galaxy  & axisymmetric & 3D & 1 million   \\ \hline
3D     & random   & axisymmetric & 3D & 0.8 million \\ \hline
2D    & random in xz plane   & axisymmetric & 2D & 0.8 million  \\ \hline
Galaxy-sp & spherical galaxy & spherical    & 3D & 1 million\\ \hline
3D-sp     & random    & spherical    & 3D & 0.8 million \\ \hline
2D-sp     & random in xz plane  & spherical    & 2D & 0.8 million \\ \hline
\end{tabular}\label{Table 1}

\end{table*}

\section{Results}

\subsection{Prominent orbit families}

\begin{figure*}[]
\centering
\includegraphics[width=50mm]{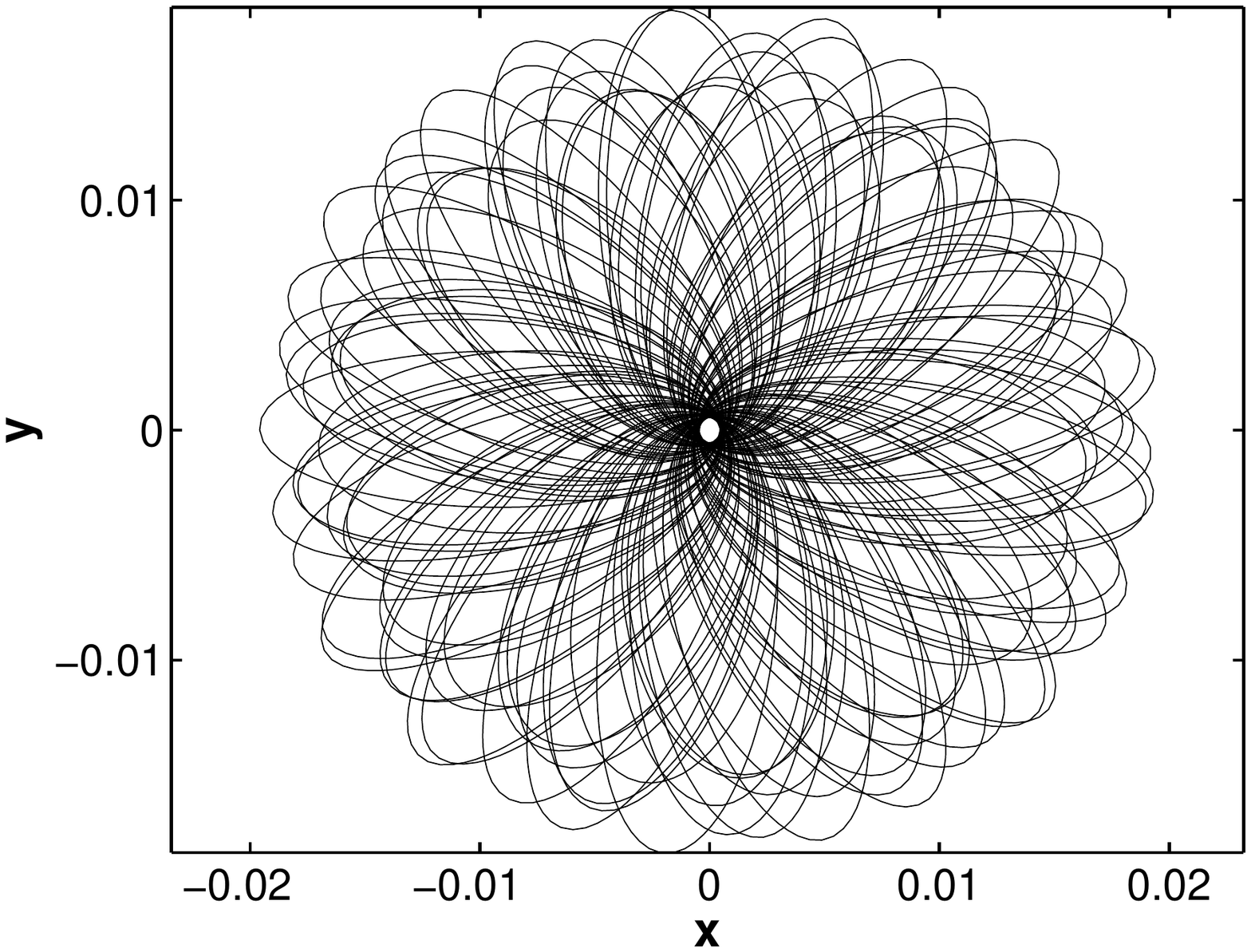}
\includegraphics[width=50mm]{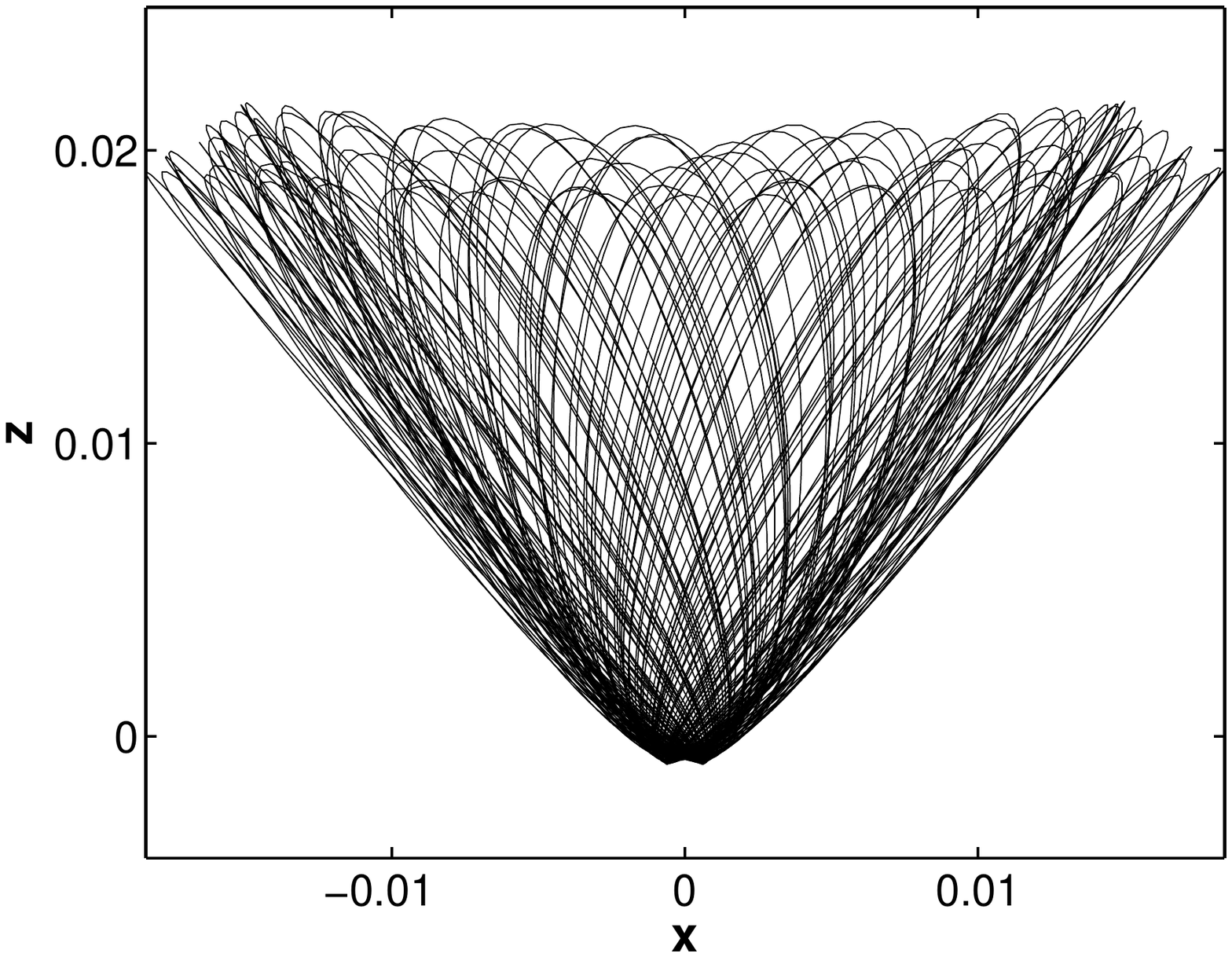}
\includegraphics[width=50mm]{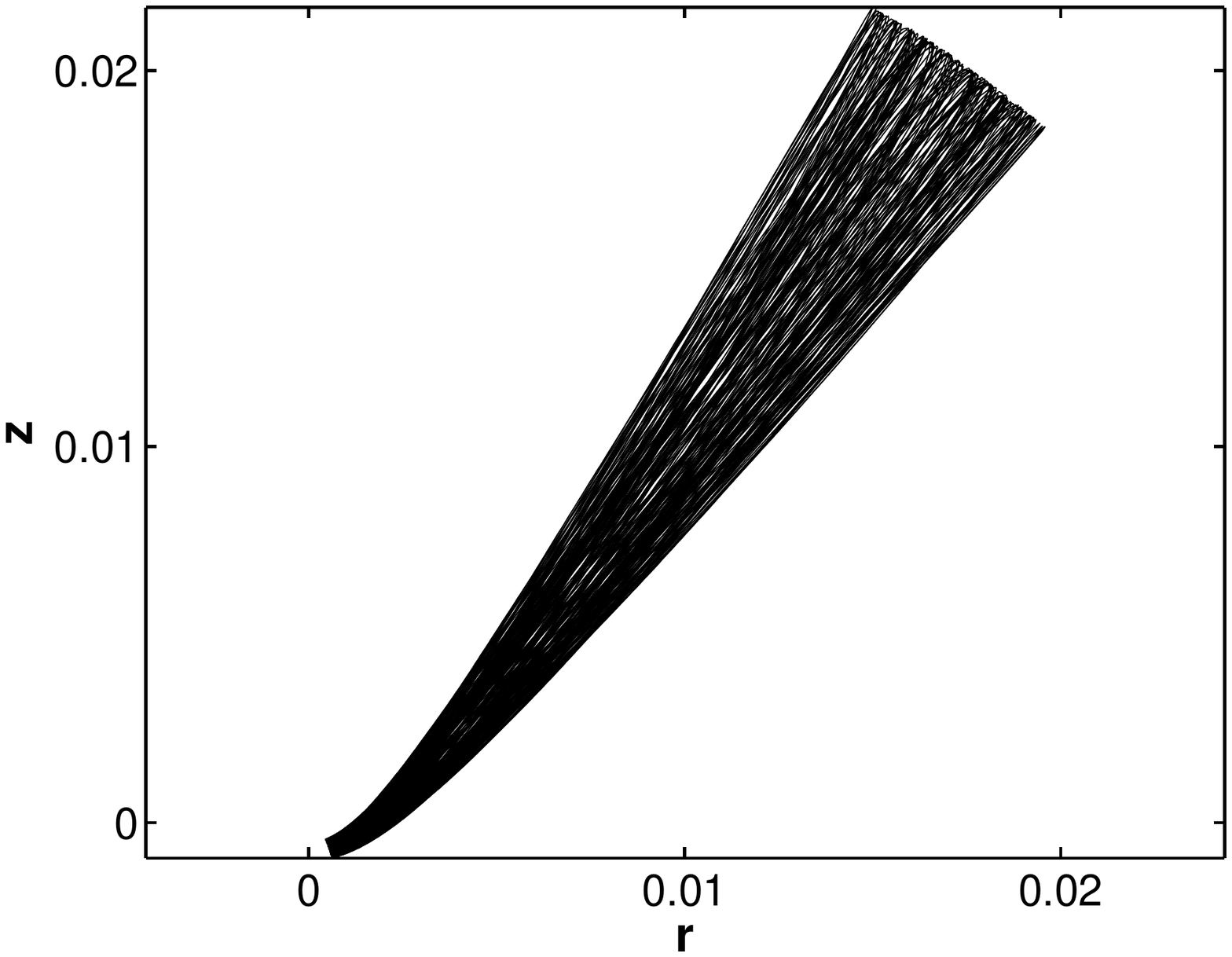}
\caption{A typical saucer orbit that emerged in the Galaxy and 3D simulations. The three panels from left to right show the orbit projection in the x-y, x-z and r-z plane respectively, where $r=\sqrt{x^2+y^2}$.}
\label{saucer}
\end{figure*}

In theory, axisymmetric potentials are thought to harbor resonant, centrophilic orbits \citep{V14} that bear some broad similarity to those in triaxial systems \citep{ST99, SS00}; the main difference is that since the degree of symmetry is higher in axisymmetric systems, they should admit fewer chaotic orbits and, naturally, more 1:1 resonances within the symmetry plane \citep{PM01}. Of particular note in an axisymmetric model is the saucer orbit, predicted within an analytical potential \citep{R82, LS92}. We identified saucer orbits within our N-body model of an axisymmetric galaxy, even though the potential is neither a homologous figure nor an analytic potential. Figure \ref{saucer} shows the projection on x-y plane, x-z plane and R-z plane of the saucer orbit within our 3D run. This orbit traversed the inner 0.7 parsec of the model, with pericenter passes only 0.05 parsec from the SMBH. These orbits are thought to be key in interacting with and being scattered by binary SMBHs.

\begin{figure*}[]
\centering
\includegraphics[width=50mm]{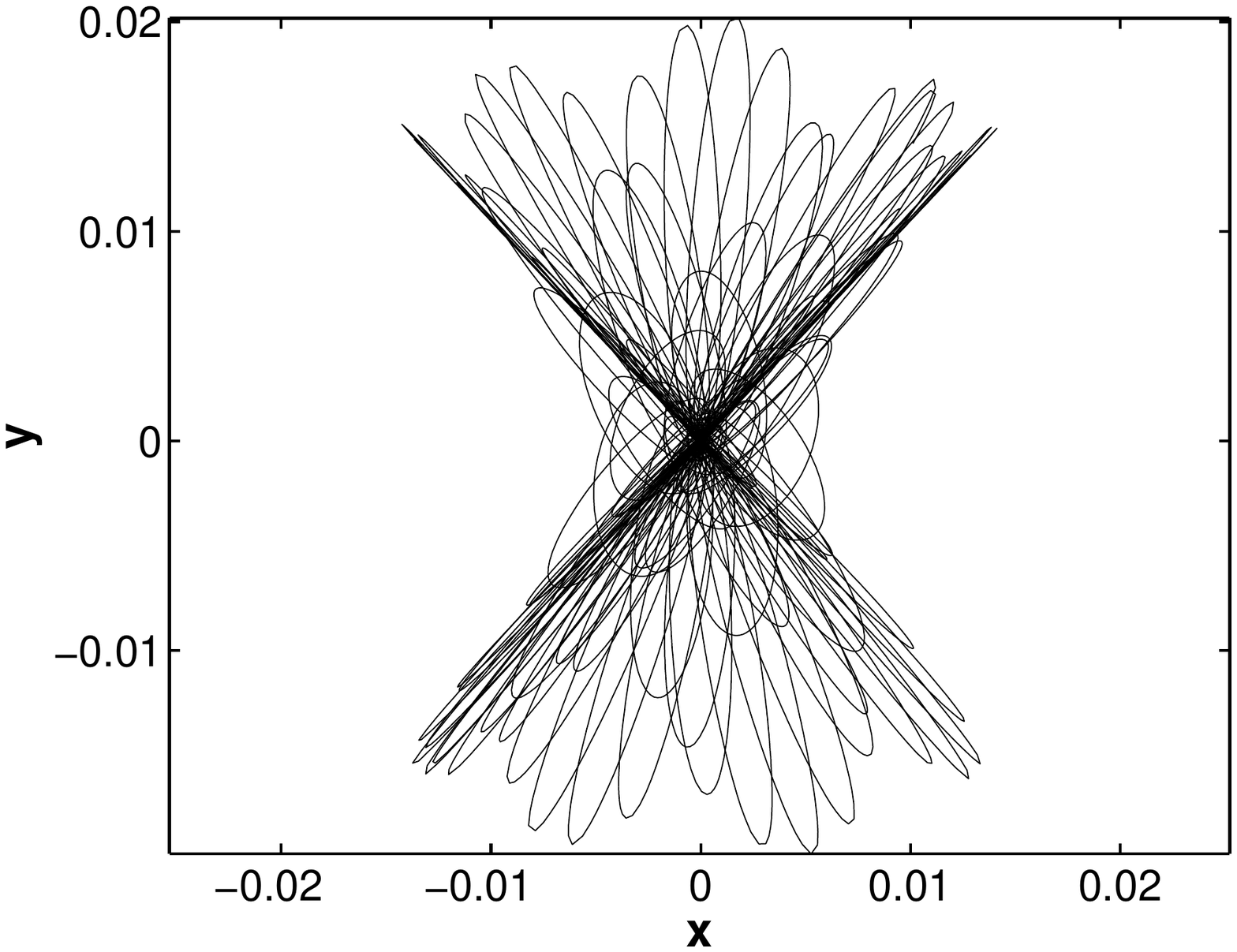}
\includegraphics[width=50mm]{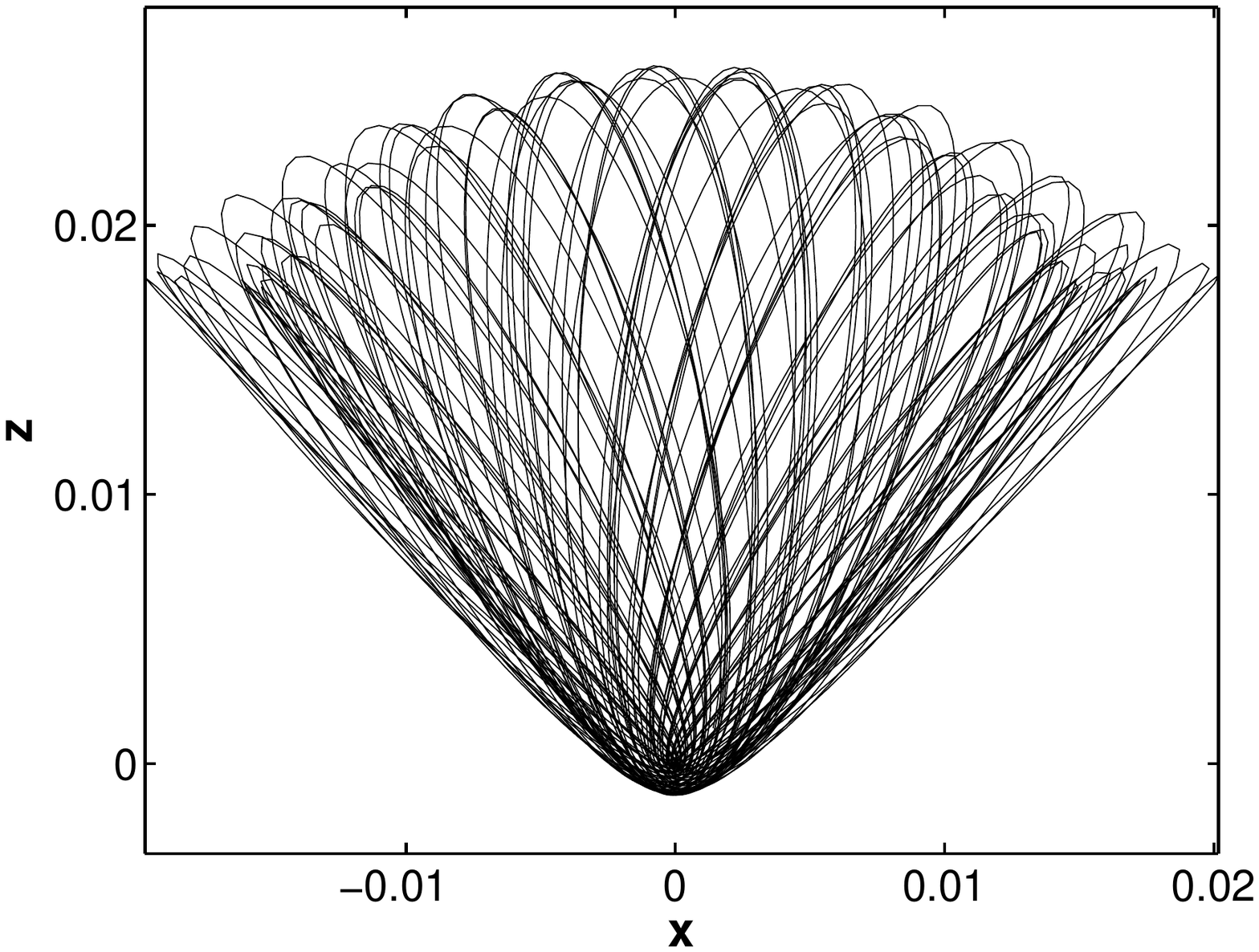}
\includegraphics[width=50mm]{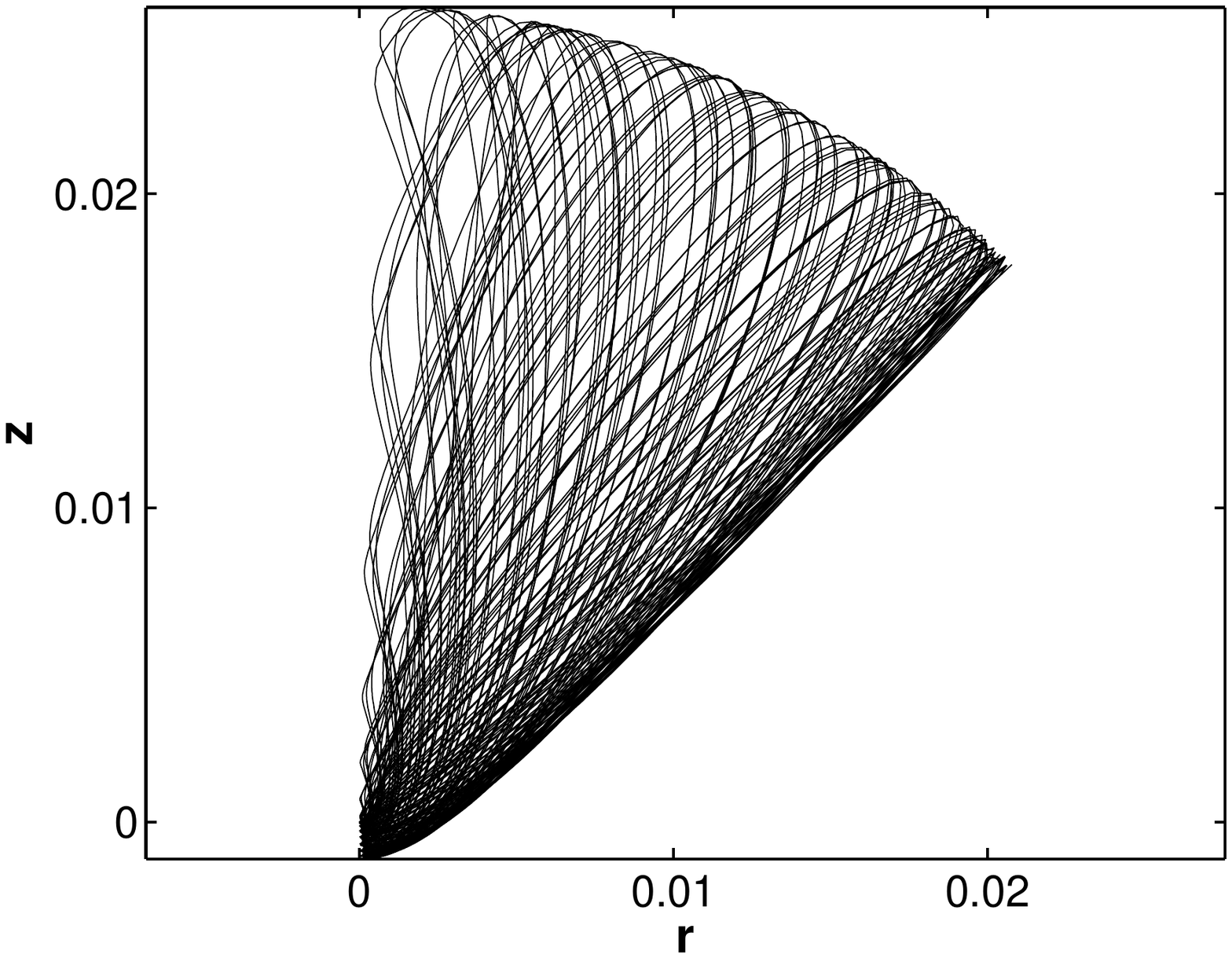}
\caption{A typical pyramid orbit emerged in the Galaxy and 3D simulations. The three panels from left to right show the projection of the pyramid orbit in the x-y, x-z and r-z plane respectively, where $r=\sqrt{x^2+y^2}$. Note the hole in the x-y plane projection of the saucer orbit, which is not present in the pyramid orbit.}
\label{pyramid}
\end{figure*}

Unexpectedly we also found pyramid orbits\citep{ST97, MV99, ST99, PM01}, which are thought to exit only in triaxial galaxies, as they originate from breaking the symmetry axis of a saucer parent orbit \citep{MV99}. Figure \ref{pyramid} displays a pyramid orbit from our simulation. From the x-y plane projection, it is clear that the pyramid passes through the center of the galaxy, while the saucer does not. These are also ideal orbits to comprise the loss cone for binary black hole coalescence. In our model, these pyramid orbits only exist in the part of the model that exhibits slight triaxiality within 2  parsecs of the SMBH. With such a minor degree of triaxiality, it is perhaps surprising that these orbits exist at all; indeed, it is not clear how small the deviation from non-axisymmetry must be to admit these formally triaxial orbits.

We found these distinctive orbits in all our axisymmetric runs although note that in two-dimensions, pyramids and saucers do not distinguish from each other \citep{MV11}. Through the observation of hundreds of orbits, we defined the criteria to separate saucers from pyramids within our Galaxy run, where the orbits are directly from the N-body model. Saucers satisfy $-2.2<E<-1.7$ , $fx/fz <1$, $fy/fz<1$, and  $1.74\times 10^{-4}<L_{min}<0.0035$. In our particular potential, pyramids satisfy $-2.2<E<-1.7$ , $fx/fz <1$, $fy/fz<1$, and $L_{min}<=1.75\times 10^{-4}$. There are approximately 600 saucers and 150 pyramids in the Galaxy run.

Since we are motivated to examine orbits to better understand how they promote rapid SMBH coalescence, we search for ``orbits of interest" within our Galaxy model\citep{V14}. These orbits could potentially lie within the binary SMBH loss cone, and are a composite of formally centrophilic orbit families such as boxes or pyramids, as well as those orbits with pericenters roughly that of the separation between SMBHs in \citet{KHBJ13}, including chaotic orbits. In our axisymmetric model, there are over 14000 such orbits, with a total mass of 0.014 in system units, which is 3 times larger than the SMBH, while the spherical model only hosts about 2000 of these orbits.

\begin{figure}[]
\centering
\includegraphics[width=87mm]{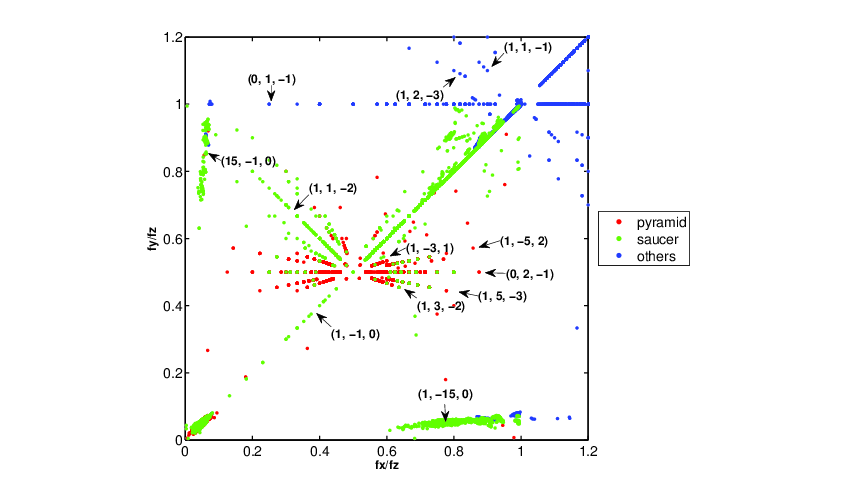}
\caption{fy/fz versus fx/fz for 0.1 million particles with E=-2 in the 3D simulation. Pyramids are denoted by red dots, saucers by green dots and others by blue dots. Both saucers and pyramids have $fy/fz<1$ and $fx/fz<1$ and small angular momentum. The angular momentum of saucers are $1.75\times 10^{-4}<L_{min}<0.0035$, while that of the pyramids are $L_{min}<=1.75\times 10^{-4}$. In this energy slice, the saucers are $15\%$ and the pyramids are $6\%$. Since these are resonant orbits, they will lie on a distinct line in this frequency map. We mark notable resonance lines by (u,v,w), which correspond to integers and are coefficients of the equation $u\cdot fx+v\cdot fy+w\cdot fz=0$. Pyramids mainly lie on lines (1, -3, 1) (1, -5, 2), (0, 2, -1), (1, 5, -3) and (1, 3, -2), while saucers are present on (15, -1, 0), (1, -15, 0), (1, 1, -2) and (1,-1,0). }
\label{fft}
\end{figure}

In our 3D simulation, where we can more finely-sample the orbit content based on the initial energy of the orbit,  saucers and pyramids are primarily evident in the deep energy slice at E=-2. In this region, they are also separately distributed in frequency and angular momentum space. 
Figure \ref{fft} shows fy/fz versus fx/fz. The red dots are pyramids, green ones are saucers, others are in blue. We can easily see from this figure that the saucers mainly lie on the fx=fy diagonal line, while pyramids spread around the line fy/fz=0.5. Saucers and pyramids are also easily separable in angular momentum at this fixed energy slice; saucer orbits comprise 15 percent of the total mass of this energy slice, while pyramids are 6 percent.

\begin{figure}
\centering
\includegraphics[width=75mm]{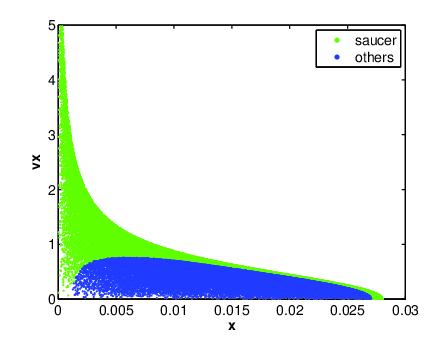}
 \caption{Surface of section of vx versus x of E=-2 slice in the 2D simulations. The saucers are in green and others in blue. Saucers have the angular momentum $L_{min}< 0.0035$. It is seen that in the upper part of the figure, the angular momentum of the particles are smaller. }
\label{sosE10}
\end{figure}

Figure \ref{sosE10} shows the surface of section of the E=-2 slice in the 2D simulation, the green dots are saucers, the blue ones are others. Saucers are those with a minimum angular momentum less than 0.0035, while other orbits have larger angular momenta, and recall that in 2D simulations, saucers and pyramids are the same \citep{MV11}.

\subsection{Global orbital structure}

Though we concentrated on the orbits that could encounter and interact with a binary SMBH in each model, there are a rich variety of resonant orbits throughout the system, and we discuss the global orbital content here.

\begin{figure*}
\centering
\includegraphics[width=80mm]{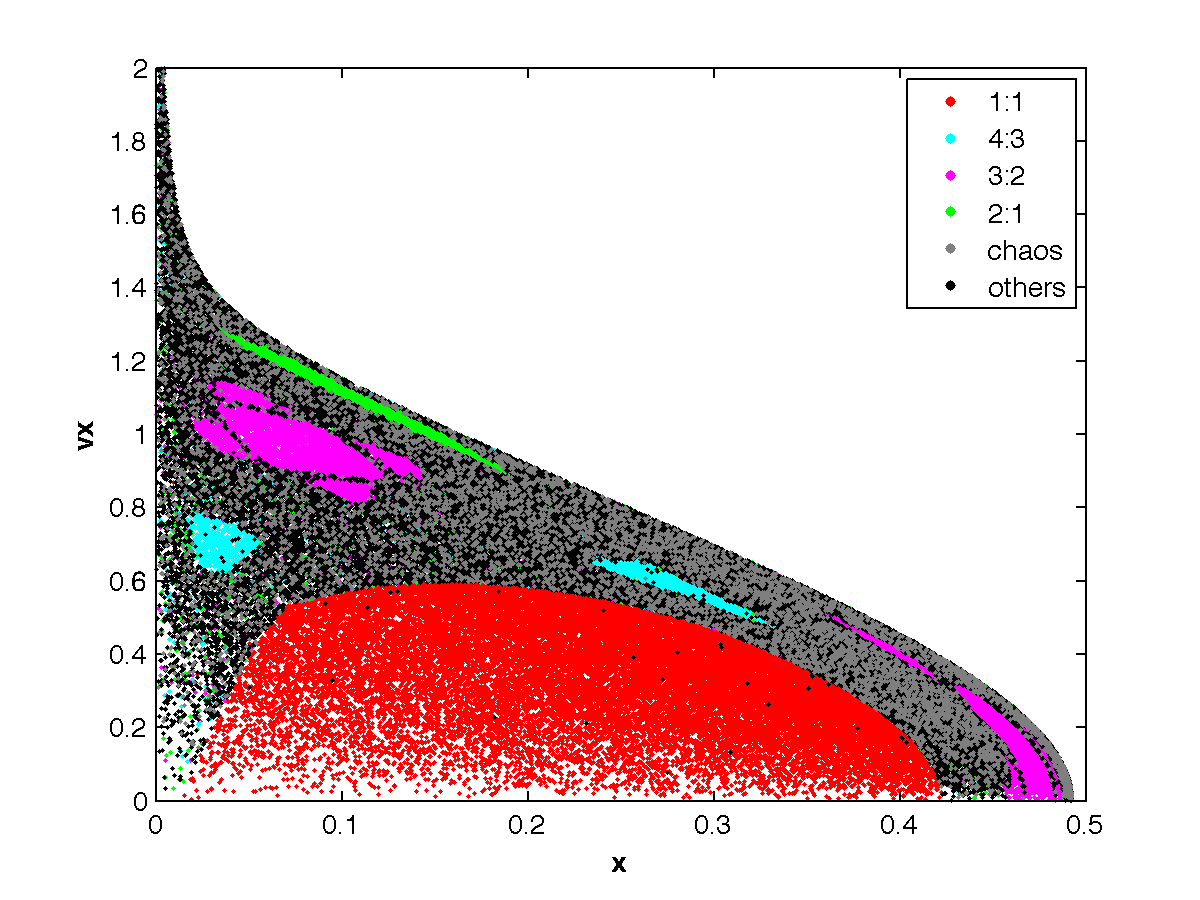}
\includegraphics[width=80mm]{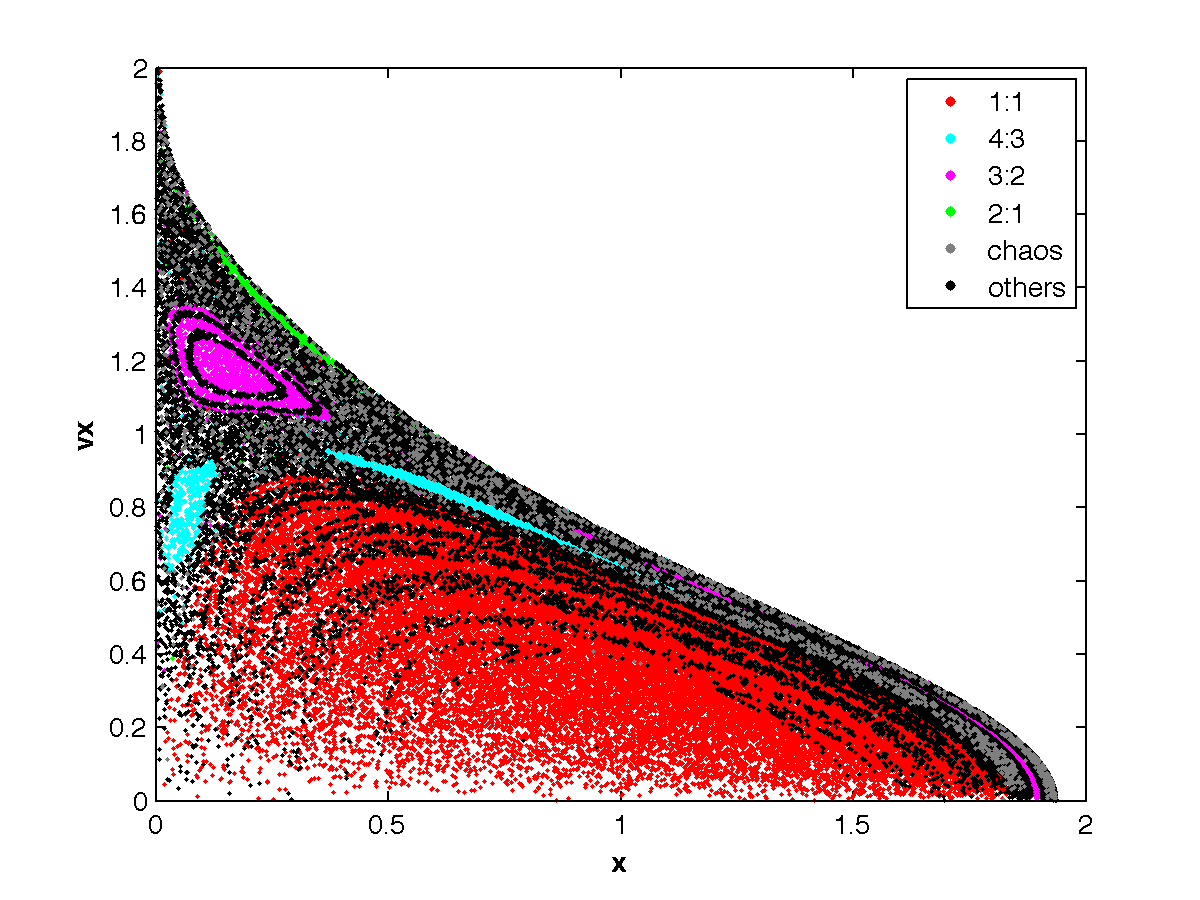}
\caption{Surface of section in the 2D simulations. The two panels from left to right show the surface of section of E=-1,-0.4, slice respectively in the ax-2D-random simulations. The stellar mass of particles with energy less than each E in the axisymmetric galaxy is respectively $10\%$ and $45\%$ of the total stellar mass. The dots are colored by fz/fx, in which 1:1 loops are denoted by red dots, 4:3 pretzels by cyan dots, 3:2 fishes by magenta dots, 2:1 bananas by green dots, chaos by grey dots and other resonances by black dots. The 1:1 loop is always dominant. Chaotic orbits always occupy the lower angular momentum part of the figure, as they interact with the SMBH.}
\label{sos}
\end{figure*}

Figure \ref{sos} shows the surface of section of two energy slices in the 2D run, colored by the fz/fx ratio to denote different orbit families. It is readily apparent from the large area occupied by the 1:1 loop orbit that it is the dominant family; in the spherical model it is the only regular, non-chaotic, orbit family.
Fish, pretzels also feature significantly in these energy slices, and though the fraction of chaotic orbits are small, they are present peppered throughout the region occupied by high-order resonances.

\begin{figure}
\centering
\includegraphics[width=75mm]{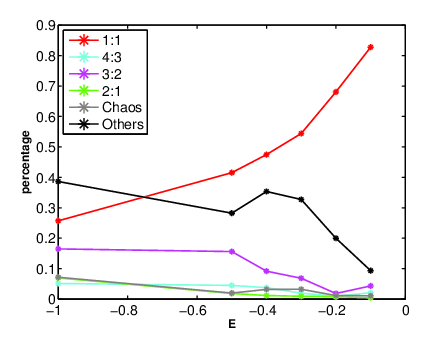}
\caption{The percentage of different type of orbits as a function of energy in 2D simulations. This plot shows the percentage of each type of orbits presented in Figure \ref{sos} with the same legend. The trend is the rate of 1:1 loop keeps increasing as the energy rises, while the rates of nearly all the other types decrease, of which only the 3:2 fishes and ``other resonances" are ever over $10\%$.}
\label{pEax2d}
\end{figure}

Figure \ref{pEax2d} presents the percentage of different orbit types as a function of energy. It can be seen that 1:1 loop orbits are the dominant orbit family at nearly every energy; ``other resonant" orbits begin to dominate only at the slice that is most highly-bound to the SMBH. At the least-bound energy slice the percentage of 1:1 loops is higher than 85$\%$ and this may be partially due to the fact that this slice contains some orbits near the physical outskirts of the system, where adiabatic squeezing is less effective at transforming the shape. The fraction of low-order resonant orbits increases for  more tightly-bound orbits. Aside from the loop orbit, the 3:2 fish orbit family is the most prominent of the ones we tracked. The percentages of 4:3 pretzels, 2:1 bananas higher-order resonances and chaotic orbits are always below 10$\%$.  

\begin{figure*}[]
\centering
\includegraphics[width=85mm]{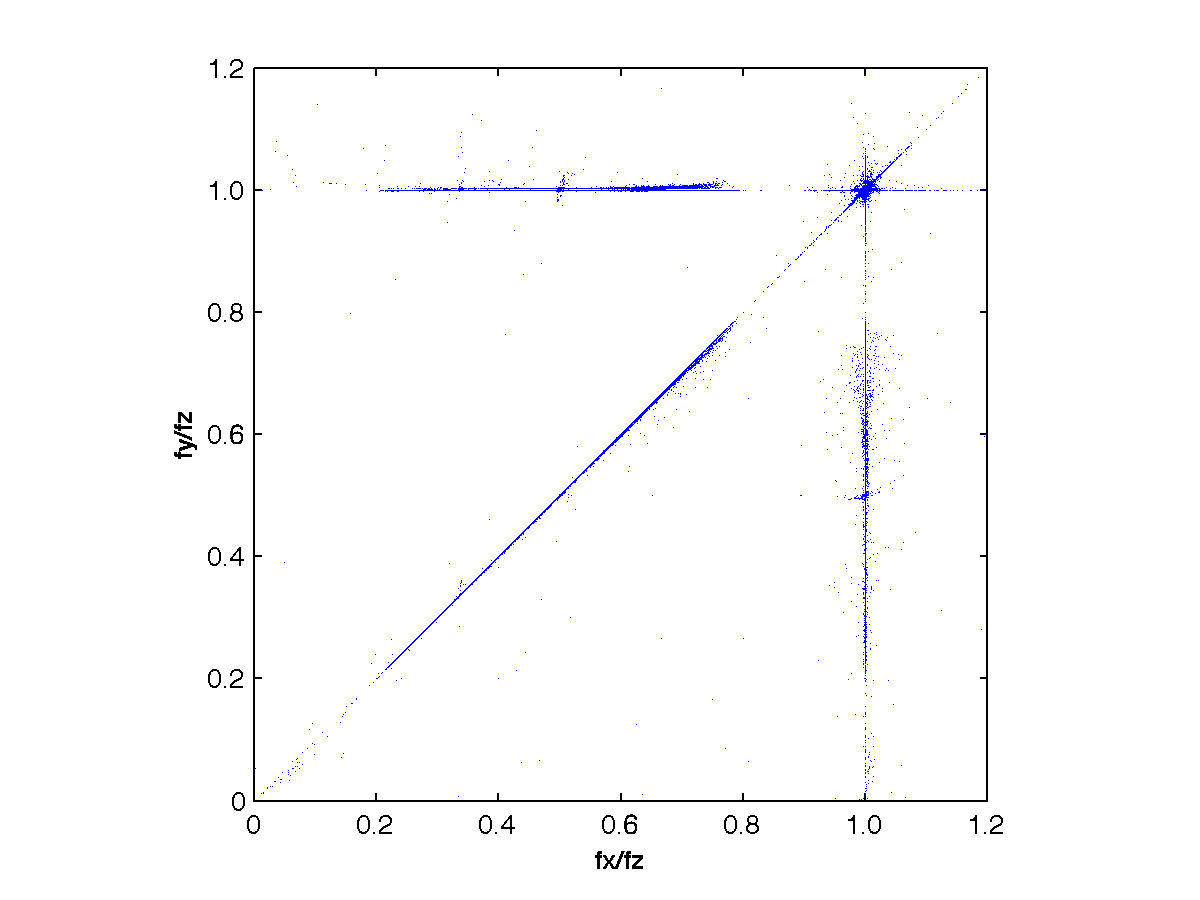}
\includegraphics[width=85mm]{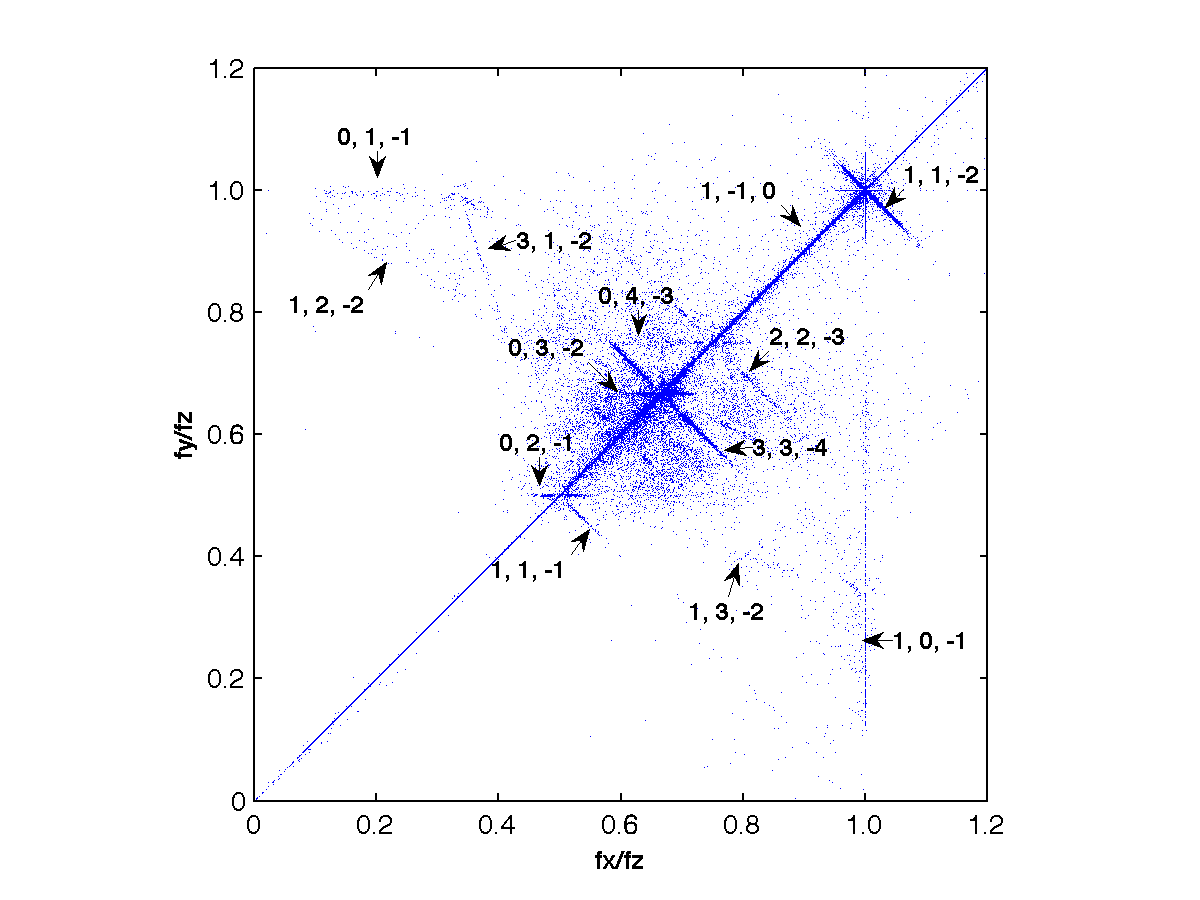}
\caption{Left: fy/fz versus fx/fz for Galaxy-sp simulations. $88\%$ particles are lying at the point (1,1), which means they have fx=fy=fz and are 1:1:1 tubes, and nearly all the rest particles reside on lines fx=fy, fx=fz and fy=fz; there are no complex orbit types in this model. Right: fy/fz versus fx/fz for Galaxy simulations. In contrast to the Galaxy-sp model, the Galaxy model has a rich variety of orbits such as (1, 1, -2) and (3, 3, -4), (0, 3, -2), (0, 2, -1) and (1,1,-1), etc. However $98\%$ particles are still short-axis tubes.}
\label{fftscf}
\end{figure*}

The left and right panels of Figure \ref{fftscf} show fy/fz vs fx/fz of the Galaxy-sp and Galaxy simulation respectively. In the spherical galaxy model, 88$\%$ particles lie around the (1,1) point, which means they have the fx:fy:fz=1:1:1, while in the axisymmetric model this percentage is 33$\%$. However in the axisymmetric model the percentage of particles lying on the line fx=fy is 98$\%$; these are short-axis tubes. The resonance orbits lying on line $u\cdot fx+v\cdot fy+w\cdot fz=0$ in the axisymmetric model are also marked by the line coefficient (u,v,w) as showed in the figure. It is seen that comparing with the Galaxy-sp model showed in the left panel, the Galaxy model has a rich variety of orbits such as (1, 1, -2), (3, 3, -4), (0, 3, -2), (0, 2, -1) and (1,1,-1), etc..

\begin{figure*}[]
\centering
\includegraphics[width=78mm]{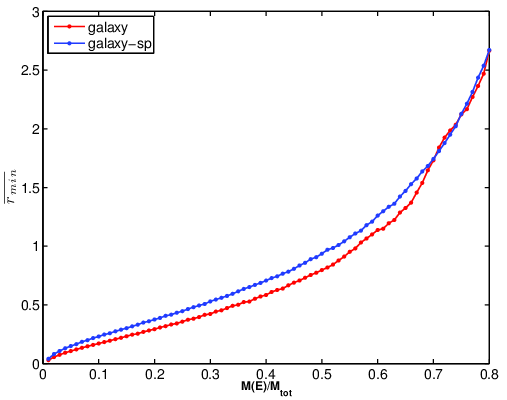}
\includegraphics[width=72mm]{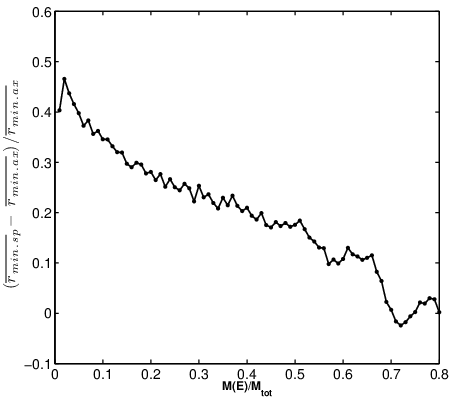}
\caption{The left panel shows $\overline{r_{min}}$, the average pericentric distance in each mass bin, as a function of mass fraction. The blue line is for Galaxy-sp model and the red line for Galaxy model. The right panel quantifies the difference between the pericentric distance in each model: $\left(\overline{r_{min,sp}}\; -\: \overline{r_{min,ax}}\;\right)/\overline{r_{min,ax}}$, as a function of mass fraction. Inside around $70\%$ mass fraction, this difference is always over $10\%$, reaching nearly $50\%$ at mass fraction of $2\%$. }
\label{rmin}
\end{figure*}

The left panel of Figure \ref{rmin} displays the mean pericenter distance of the particles in each bin as a function of mass fraction for the Galaxy (red) and Galaxy-sp (blue) run. There are 100 bins in each simulation, with 10000 particles per bin.
It is clear that the mean pericentric distances are smaller in the axisymmetric galaxy out to an enclosed mass of 70$\%$, and at that point the model is more nearly spherical. Note that we calculate the pericentric distance in ellipsoidal coordinates so that we are not biased by the more compact vertical dimension in the flattened model; in other words, $r_{min}=\sqrt{(x/a)^2 + (y/b)^2 +(z/c)^2}$. To quantify the difference between the pericentric distances more explicitly, the right panel is the difference between the Galaxy and Galaxy-sp models, weighted by the axisymmetric model. Orbits delve $50\%$ deeper into the center at mass fraction of $2\%$. For the  most part, the difference is over 10$\%$.

\section{Conclusion and Discussion}

We identified several major orbits families in our axisymmetric galaxy model such as saucers, bananas, fishes, and short-axis tubes. These orbits are present despite the relatively minor flattening (c/a=0.75) compared to a spherical model. Due to a very slight deviation from an oblate spheroid at the center of the axisymmetric model (T=0.28), pyramid orbits are also present, making up 6$\%$ of mass within the inner 0.5 parsec. It is not clear how much a system needs to deviate from axisymmetry to generate pyramids. 

Since we are primarily interested in whether the orbital content in the axisymmetric model is sufficient to drive binary black holes to coalesce, we took a census of those particles that would reside in the loss cone of a binary black hole. The total mass of particles with orbits that could interact with a hard binary black hole in the axisymmetric galaxy simulation is roughly three times that of the SMBH, and about seven times of that in the spherical galaxy simulation.  According to three-body scattering experiments, the SMBH binary needs to scatter $1.2\sim 1.5$ times its mass to transition to the gravitational wave regime \citep{QH97, SHM07}, and this is consistent with the mass of stars on potential loss cone orbits in our axisymmetric model. In a separate work, we will track which of these orbits are actually scattered by the SMBH binary as the system evolves, but it appears that the orbital content in our axisymmetric model is more than enough to drive the SMBHs to merge in less than a Hubble time.

There may be several reasons why the hardening rates in \citet{VAM14} and \citet{KHBJ13} differ. A suggestion has been made that numerical relaxation may have artificially 
enhanced SMBH binary scattering in \citet{KHBJ13}, while another idea posed is that the system in \citet{KHBJ13} is more perturbed from virial equilibrium, and it is this time-dependent perturbation that refills the loss cone (R. Spurzem, private communication). The results of our work imply that the slight triaxiality in our model inside the radius of influence of the black hole may be the key in explaining the apparent difference between the two results. The triaxial center in our model increased the number of potential loss cone orbits near the black hole, spawning formally centrophilic orbit families -- like pyramids -- to appear, and allowing for a wide diffusion of orbits in angular momentum. If it is true that the central shape is a major factor in the differing SMBH coalescence times in these two papers, we are left with several interesting and related questions: what is the orbital content for more realistically-flattened models?; how small a deviation from pure axisymmetry is required to gain enough centrophilic orbits to drive SMBH binaries to coalesce?; and are any real galaxies perfectly axisymmetric enough to pose a final parsec problem?

\acknowledgements

We thank Rainer Spurzem for discussions. B.L. acknowledges support from Vanderbilt Discovery Grant. K.H.-B. acknowledges support from NSF CAREER award AST-0847696, as well as the support from the Aspen Center for Physics. The simulations were performed on the dedicated GPU cluster ACCRE at the Advanced Computing Center for Research and Education at Vanderbilt University, Nashville, TN, USA.

\end{document}